\documentclass[twocolumn,twoside,slac_two]{revtex4}
\usepackage{graphicx}
\usepackage{fancyhdr}
\begin{document}

%
\def \ie    {\hbox{\it i.e.}}     
\def \etc   {\hbox{\it etc.}}
\def \ibid  {\hbox{\it ibid.}}
\def \vs    {\hbox{\it vs.}}
\def \eg    {\hbox{\it e.g.}}     
\def \cf    {\hbox{\it cf.}}
\def \etal  {\hbox{\it et al.}}
\def \via   {\hbox{\it via}}
\def \CPC   {Comput. Phys. Commun.~}
\def \EPJ   {European Phys. Journal~}
\def \JHEP  {J. High Energy Phys.~}
\def \NIM   {Nucl. Instr. Meth.~}
\def \NP    {Nucl. Phys.~}
\def \PL    {Phys. Lett.~}
\def \PRD   {Phys. Rev. D~}
\def \PRL   {Phys. Rev. Lett.~}
\def \PRep  {Phys. Reports~}
\def \RMP   {Rev. Mod. Phys.~}
\def \zphys {Z. Phys.~}
\hyphenation{back-ground}
\hyphenation{brem-sstrah-lung}
\hyphenation{cal-or-ime-ter cal-or-ime-try}
\hyphenation{had-ron had-ronic}
\hyphenation{like-li-hood}
\hyphenation{posi-tron posi-trons}
\hyphenation{semi-lep-tonic}
\hyphenation{syn-chro-tron}
\hyphenation{system-atic}
%
%
\def \beq   {\begin{equation}}
\def \eeq   {\end{equation}}
\def \bbeq  {\begin{eqnarray*}}
\def \ebeq  {\end{eqnarray*}}
\def \bbeqn  {\begin{eqnarray}}
\def \ebeqn  {\end{eqnarray}}
\def \Tr    {\mathop{\mathrm Tr}}
\def \Im    {\mathop{\mathrm Im}}
\def \Re    {\mathop{\mathrm Re}}
\def \vect  {\overrightarrow}
\def \twdl  {\widetilde}
\def \hat   {\widehat}
\def \partder#1#2  {\partial #1\over\partial #2}
\def \secder#1#2#3 {\partial^2 #1\over\partial #2 \partial #3}
%
%
\def \omg#1 {\mbox {${\mathcal O}(#1)$}}
\def \avg#1 {$\left\langle #1\right\rangle$}
\def \to    {\rightarrow}
\def \bra#1 {$\left\langle #1\right|$}
\def \ket#1 {$\left| #1\right\rangle$}
\def \braket#1#2 {\left\langle #1\right. \left| #2\right\rangle}
\def \amp#1 {${\mathcalA}(#1)$}
\def \apgt  {\raisebox{-0.6ex}{$\stackrel{>}{\sim}$}}
\def \aplt  {\raisebox{-0.6ex}{$\stackrel{<}{\sim}$}}
\def \pma#1#2 {\mbox{\raisebox{-0.6ex}
           {$\stackrel{\scriptstyle \;+\; #1}{\scriptstyle \;-\; #2}$}}}
%
\def \dr    {$\;^\mid\!\!\!\longrightarrow$}
%
\def \ev    {\,\mathrm {eV}}
\def \kev   {\,\mathrm {keV}}
\def \mev   {\,\mathrm {MeV}}
\def \gev   {\,\mathrm {GeV}}
\def \tev   {\,\mathrm {TeV}}
\def \km    {\,\mathrm {km}}
\def \cm    {\,\mathrm {cm}}
\def \mm    {\,\mathrm {mm}}
\def \um    {\,\mu\mathrm m}
\def \ghz   {\,\mathrm {GHz}}
\def \mhz   {\,\mathrm {MHz}}
\def \khz   {\,\mathrm {kHz}}
\def \ps    {\,\mathrm {ps}}
\def \ns    {\,\mathrm {ns}}
\def \us    {\,\mu\mathrm s}
\def \ms    {\,\mathrm {ms}}
\def \hz    {\,\mathrm {Hz}}
\def \pb    {\,\mathrm {pb}^{-1}}
\def \fb    {\,\mathrm {fb}^{-1}}
\def \mrad  {\,\mathrm {mrad}}
\def \BR#1#2 {\mbox{Br}(#1$\to$#2)}
\def \JP     {\mathrm J$^{\mathrm P}$}
\def \Mw    {${\mathrm M}_W$}
\def \Mz    {${\mathrm M}_Z$}
\def \Mpi   {${\mathrm M}_\pi$}
\def \Mk    {${\mathrm M}_K$}
\def \Gf    {G$_{\mathrm F}$}
\def \As    {$\alpha_s$}
\def \Mt    {${\mathrm M}_{t}$}
\def \Mb    {${\mathrm M}_{b}$}
\def \Mc    {${\mathrm M}_{c}$}
\def \Ms    {${\mathrm M}_{s}$}
\def \Mud   {${\mathrm M}_{u,d}$}
\def \Mh    {${\mathrm M}_{H}$}
\def \sW    {$\sin^2\theta_{\mathrm W}$}
\def \sWeff {$\sin^2\theta_{\mathrm W}^{eff}$}
\def \gV    {g$_V$}
\def \gA    {g$_A$}
\def \lms   {\Lambda_{\overline{\mathrm MS}}}
\def \Vub   {$|$V$_{\mathrm{ub}}|$}
\def \Vus   {$|$V$_{\mathrm{us}}|$}
\def \Vcb   {$|$V$_{\mathrm{cb}}|$}
\def \MET   {\mbox{$E_T\hspace{-1.2em}\slash\hspace{1.0em}$}}
\def \PET   {\mbox{$p_T$}}
\def \pT    {\mbox{$p_\perp}}
\def \dedx {d{\it E}/d{\it x}}
\def \rphi {$r$-$\phi$}
\def \Pt2  {${\mathrm P}_\perp^2$}
%
\def \epem {$e^+e^-$}
\def \mpmm {$\mu^+\mu^-$}
\def \tptm {$\tau^+\tau^-$}
\def \ppb  {$p\overline p$}
\def \ttb  {$t\overline t$}
\def \lplm {$\ell^+ \ell^- $}
\def \J    {$\mathrm J/\psi$}
\def \Ks   {$\mathrm K^0_{\mathrm S}$}
\def \Kl   {$\mathrm K^0_{\mathrm L}$}
\def \Bs   {$\mathrm B_{\mathrm S}$}
\def \Bo   {$\mathrm B^0$}
\def \Bp   {$\mathrm B^+$}
\def \Ds   {$\mathrm D_{\mathrm S}$}
\def \Do   {$\mathrm D^0$}
\def \Dp   {$\mathrm D^+$}
\def \Bss  {$\mathrm B_{\mathrm S}^*$}
\def \Bso  {$\mathrm B^{*0}$}
\def \Bsp  {$\mathrm B^{*+}$}
\def \Dss  {$\mathrm D_{\mathrm S}^*$}
\def \Dso  {$\mathrm D^{*0}$}
\def \Dsp  {$\mathrm D^{*+}$}

\def \Klsm {\Kl $\to \pi^0 \pi^0 \mu^+\mu^-$}
\def \Klnp {\Kl $\to \pi \pi X^0, X^0 \to \mu^+\mu^-$}


\title{Search for \Klsm~with KTeV Data}

%

\author{Leo Bellantoni for the KTeV Collaboration}
\affiliation{Fermi National Accelerator Laboratory, Batavia, IL 60510, USA}

\begin{abstract}
This presentation reports on the first experimental search for the decay
\Kl $\to \pi^0 \pi^0$ \mpmm~based on data collected by the KTeV experiment.
Although this decay mode is possible within the standard model, its rate is
phase space suppressed.  The HyperCP experiment has
recently observed 3 $\Sigma^+ \to p$ \mpmm events within a narrow di-muon
mass range, suggesting that the process may occur
via a new neutral state: $\Sigma^+ \to p X^0, X^0 \to$ \mpmm with
$m(X^0) = 214.3\mev$.  The $X^0$ would create an $s$- to $d$- quark transition
at a rate that would cause \Klnp~to occur at
rates considerably over the standard model expectation.  Our preliminary results
significantly constrain this possibility.
\end{abstract}

\maketitle

\thispagestyle{fancy}

\section{Introduction							\label{sec:Intro}}
\subsection{The HyperCP Result					\label{sec:HyperCP}}

Early in 2005, the HyperCP collaboration reported~\cite{HKfound} an unusual
result in their search for the decay $\Sigma^+ \to p\mu^+\mu^-$.  They found 3
events, corresponding to a branching ratio of
[$8.6 \pma{6.6}{5.4} (stat) \pm 5.5 (syst)] \times 10^{-8}$~ for this mode.  That
value is consistent both with the expectations of the time~\cite{BrTheory1} and the
results of revisiting the calculation of the branching ratio following the
publication of the HyperCP result~\cite{BrTheory2}.  What was surprising is that
all 3 events appeared with the same reconstructed $m(\mu^+\mu^-)$ to within the
rather narrow resolution ($\approx 0.5\mev$) of the HyperCP detector.  The
probability of such a result occuring randomly as a result of only standard model
processes is less than 1\%.  Naturally, the HyperCP collaboration
speculated that there might be a contribution from a new intermediate state
with a mass of $214.3 \pm0.5\mev$.  Were that to be the case, the acceptance of
the HyperCP detector would be different than in the $\Sigma^+ \to p \mu^+\mu^-$
case, and the 3 events would correspond to
\BR{$\Sigma^+$}{$p X^0, X^0 \to \mu^+\mu^-} 
		= [3.1 \pma{2.4}{1.9} (stat) \pm 1.5 (syst)] \times 10^{-8}$.
Expressed as a partial width, $3.1 \times 10^{-8}$ corresponds to
$2.5 \times 10^{-19}\mev$.

\subsection{Response to the HyperCP Result		\label{sec:Flurry}}

The HyperCP result produced a flurry of excitement.  One early 
suggestion~\cite{GorbuRuba} was that they had observed an sgoldstino; this
interpretation suggests that \BR{$X^0$}{$\gamma\gamma$} ~might
be much greater than \BR{$X^0$}{$\gamma\gamma$} ; in response, the E391
collaboration looked for~\cite{E391} (and did not find)
\Kl $\to \pi \pi X^0, X^0 \to \gamma\gamma$, setting an
upper limit of $2.4 \times 10^{-7}$ at the 90\% C.L.

Another suggestion~\cite{HTV2007} was that HyperCP had observed the $CP-$odd $a$
Higgs boson of the next-to-minimal supersymmetric extention to the standard model
(NMSSM).  In the NMSSM, existing search techniques will not reveal the Higgs boson,
and indeed searches for the standard model Higgs have already ruled out the most
likely range of possible masses.  The minimal supersymmetric extention to the
standard model is also under some pressure in terms of theoretically possible \vs~
experimentally allowed masses for Higgs bosons, and that situation also would be
resolved by the NMSSM.  This combination of motivations lead the CLEO~\cite{CLEO},
D0~\cite{D0} and BaBar~\cite{BaBar} collaborations to search for the $a$ with
particular attention to the $m(a) = 214.3\mev$ case.  The searches all returned
null results, and the CLEO paper discusses implications of this for the NMSSM
model.

Last year, Chen \etal~suggested~\cite{Chen2008} that the HyperCP result could
be the result of a spin-1 gauge boson of the $U(1)'$ gauge model.  Of particular
significance for what follows, they found that such a model does not enhance the
rate of $K \to \pi \mu^+ \mu^-$.  This scenario was further investigated in
references~\cite{MaxU,PrinceU}.

Apart from explanations in the contexts of specific models, considerable
understanding can be gained from model-independent analyses.  If $X^0$ is
indeed a new neutral flavor changing current that creates an $s$- to $d$- quark
transition, it should appear in kaon decays.  The KTeV limit~\cite{Myself} of  
\BR{\Kl}{$\pi^0 \mu^+ \mu^-$} $< 3.8 \times 10^{-10}$ corresponds
to a partial width of $4.9 \times 10^{-24}\mev$, more than 4 orders of
magnitude below the corresponding partial width in the HyperCP result.  The
existing~\cite{PDG} world average
\BR{\Kl}{$\mu^+ \mu^-$} $= (6.84 \pm 0.11) \times 10^{-9}$ corresponds
to a partial width of $8.8 \times 10^{-21}\mev$, and the result that we present
here, \BR{\Kl}{$\pi^0\pi^0 \mu^+ \mu^-$} $< 8.63 \times 10^{-11}$ corresponds
to a partial width of $1.1 \times 10^{-24}\mev$.  Plainly, these should provide
tight constraints and indeed, more detailed analyses than these simple comparisons
of partial widths are revealing.

He \etal~in 2005 concluded~\cite{He2005} that as a consequence of the known
$K^+ \to \pi^+ \mu^+ \mu^-$ rate, the $X^0$ could not be a scalar or
vector particle.  If the HyperCP result is the result of a new pseudoscalar,
they predict (among other results) that
\mbox{Br}(\Kl $\to \pi^0 \pi^0 X^0, X^0 \to \mu^+ \mu^-$) =
										$(8.3 \pma{7.5}{6.6} ) \times 10^{-9}$.
If the HyperCP result is the result of a new axial vector, He \etal~predict
this product branching ratio would be $(1.0 \pma{0.9}{0.8} ) \times 10^{-10}$.
Deshpande \etal~\cite{Deshpande} considered spin 0 bosons and also came to the
conclusion that pseudoscalar couplings have to dominate.  Their prediction for
the pseudoscalar case branching ratio is $8.02 \times 10^{-9}$, consistent with
that of He \etal. Geng \etal~\cite{Geng2006} came to basically the same
conclusions although (as they describe in detail) there were some differences
between their calculations and the two proceeding ones.  
Chen \etal~\cite{Chen2007} discussed what the pseudoscalar and axial vector
scenarios would imply for $b$ and $\tau$ decays.  We also cite here~\cite{Oh}
the work of Oh and Tandean, which became available after this conference.

Tensor couplings evidently can not contribute~\cite{Evidently} to decays of the
type studied by HyperCP.

\subsection{Search for \Klsm~at KTeV			\label{sec:Search}}

With the KTeV detector, one could search for either
\Kl $\to \pi^+\pi^-\mu^+\mu^-$ or \Kl $\to \pi^0\pi^0\mu^+\mu^-$; in fact, the
neutral mode is the better choice.  The difference in rest masses between the
\Kl~and the final states is quite small, and the $4.6\mev$ difference between
the mass of the charged and neutral pion creates a factor of 10 difference
in available phase space.  This is the primary cause of the difference in
predicted rates for the charged-pion \vs~neutral-pion modes in
reference~\cite{He2005}.  Secondly, although the geometric acceptance of the
KTeV detector decreases as a rule with the number of particles in the final
state, the excellent energy resolution of the CsI electromagnetic calorimeter
makes $\pi^0$ detection relatively easy.

The KTeV detector has been described in detail elsewhere~\cite{KTeV}.  We will
here discuss the performance of two elements of it which are crucial for this
analysis.  The electromagnetic calorimeter was made of pure CsI, and was 27
radiation lengths deep.  For photons over $10\gev$, the energy resolution was
better than 1\%.  With electrons we obtained a positional resolution of about
$1\mm$.  In a fixed target \Kl~ experiment, where the decay point varies from
event to event, the mass resolution of a $\pi^0$ depends on the precision of
the decay point, which in turn depends on the \Kl~decay mode.  That said, 
resolutions as low as $2\mev$ in $m(\pi^0)$ have been obtained.  The muon
system was constructed of $5.1\,\mathrm m$ of steel, which is about 31 hadronic
interaction lengths.  For muons of $10\gev$ or more, the efficiency was over
98\%.  The probability for a $\pi^\pm$ to punch through the muon system and
appear as a muon was about $(1.69 + 0.17P[\mathrm {GeV}]) \times 10^{-3}$.
In the rare-decay configuration, $733 \times 10^9$ \Kl~decays occured in the
decay volume during 2 separate data taking periods begining in 1996 and 1999,
respectively.

The number of \Kl~decays in the sample is measured by counting \Kl $\to 3\pi^0$
decays, where one of the 3 $\pi^0$s decays to $e^+e^-\gamma$.  Our results
are normalized to this well-understood mode, and our reported branching ratio
limits are the result of multiplying our measurement by the branching ratios
for this normalization mode.  The normalization mode is selected to be as
similar in detection signature to the signal mode as is possible in order to
cancel systematic uncertainties.

\section{Analysis Procedure
		\label{sec:Analysis}}

The definitive description of the analysis is the thesis of David
Phillips~\cite{Dave09}.  The interested reader should also examine Dave's
contribution to the proceedings of the KAON09 conference~\cite{KAON09}.

Signal candidates are required to have 4 clusters of energy in the CsI
calorimeter which are not associated with tracks.  Two tracks of opposite
charge assignment with matching hits in the muon system and momentum over
$7\gev$ are required, and the calorimeter clusters created by the muon
candidates must have less than $1\gev$ of energy.  The 4 photons and 2 muons
must have a sum of momenta perpendicular to the \Kl~line of flight less than
$\sqrt{0.00013}$ $\gev$; this helps assure that all of the products of a single
decay have been reconstructed.  The kinematic requirement $m(\mu^+\mu^-)$
is applied, and the reconstructed $m(\pi^0)$ values have to be within $9\mev$
of the accepted value.  The reconstructed mass of all of the decay products for
signal candidates must lie within the range $495$ - $501\mev$.

There is a second signal box which examines the \mpmm~pair for consistency
with the HyperCP result; this allows us to obtain results for \Klnp~in addition
to \Klsm.  The second signal box is defined as
$213.8 \leq m(\mu\mu) \leq 214.8\mev$,
$| p_\perp^2(\mu\mu) - p_\perp^2(\pi\pi) | \leq 0.0007(\mathrm {GeV^2})$.
A single detected signal event would correspond to a branching ratio of
$3.75 \times 10^{-11}$ for \Klsm~and $4.10 \times 10^{-11}$ for \Klnp.

Background events are so-called 'accidentals', where particles from a 2nd decay that
occurs in the beam at the same time as the \Kl~decay lead to misinterpretation
of the event.  Accidental backgrounds are modeled in the simulation by
overlaying a simulated decay with an event taken from the detector on a trigger
that fires randomly.  This method is quite effective when the decay has a
relatively low branching ratio and for those cases, we have been able to
simulate event samples many times larger than the actual dataset.  None of
these events pass the selection criteria.  This method is less effective for
modes such as \Kl $\to \pi^\pm \mu^\mp \nu$, where a sample of 4.4 billion
events corresponds to only 3.2\% of the data sample.

Ultimately we need to control our background levels with an understanding
of the basic kinematics of our signal process and an examination of the rate
at which data events appear near but not in our signal boxes.  The small phase
space for the decay make background rejection relatively easy through the
requirement that the reconstructed mass of the 6 decay product candidates
match the \Kl~mass.  A small phase space means that in the \Kl~frame, the 6
decay products have low momenta; but as the accidentally coincident decaying
particle will not have the same lab-frame momentum as the \Kl, the accidental
decay products will have a considerable boost in the \Kl~frame.  As a result,
background events will tend to be higher in reconstructed \Kl~mass than the
signal.  In the end, we observe very low background rates and take the mildly
conservative approach of setting the total background rate to 0.

\section{Result and Discussion
		\label{sec:Output}}

\begin{figure*}[t]
\centering
    \scalebox{0.3}{\includegraphics{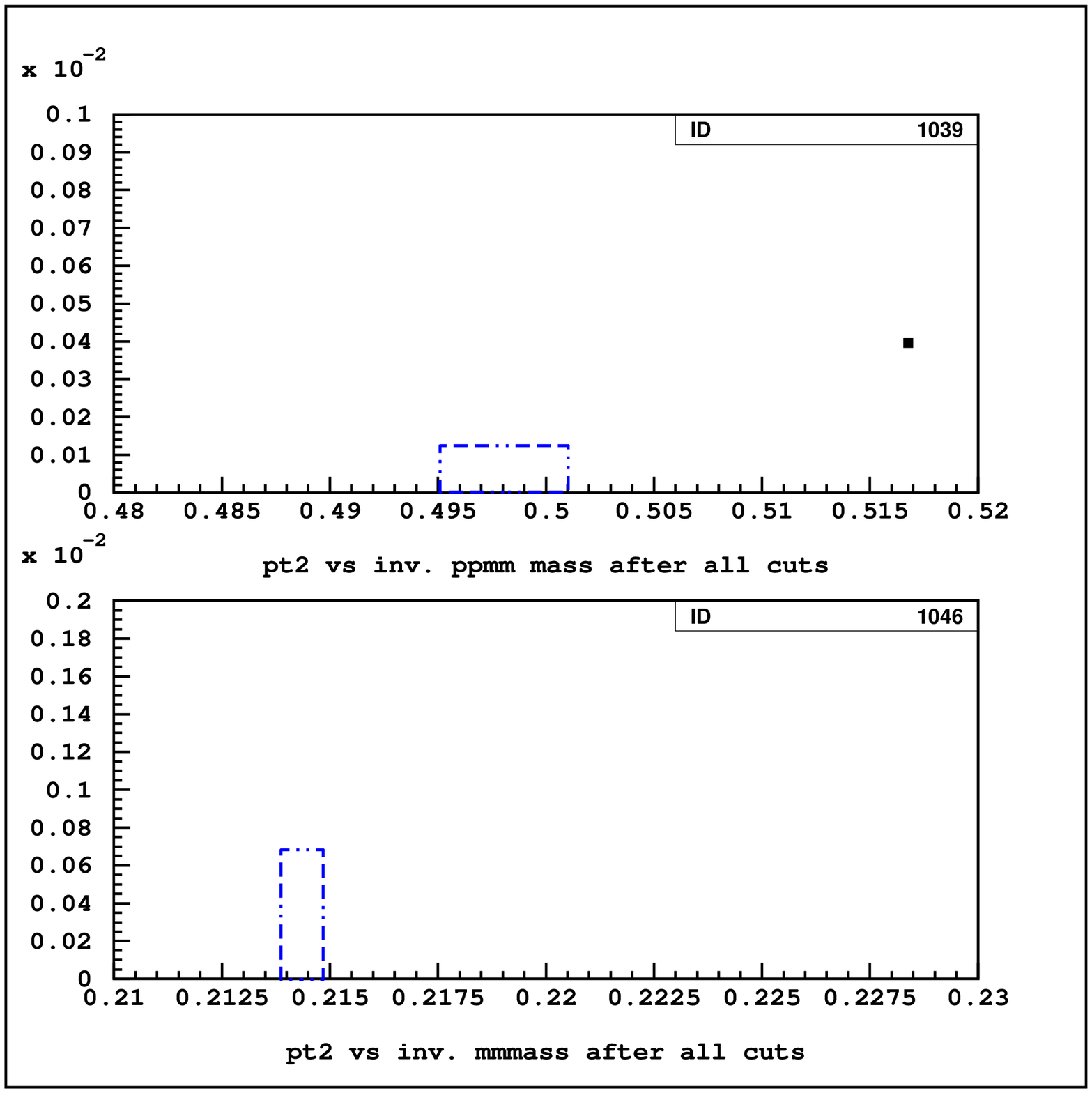}}
    \scalebox{0.3}{\includegraphics{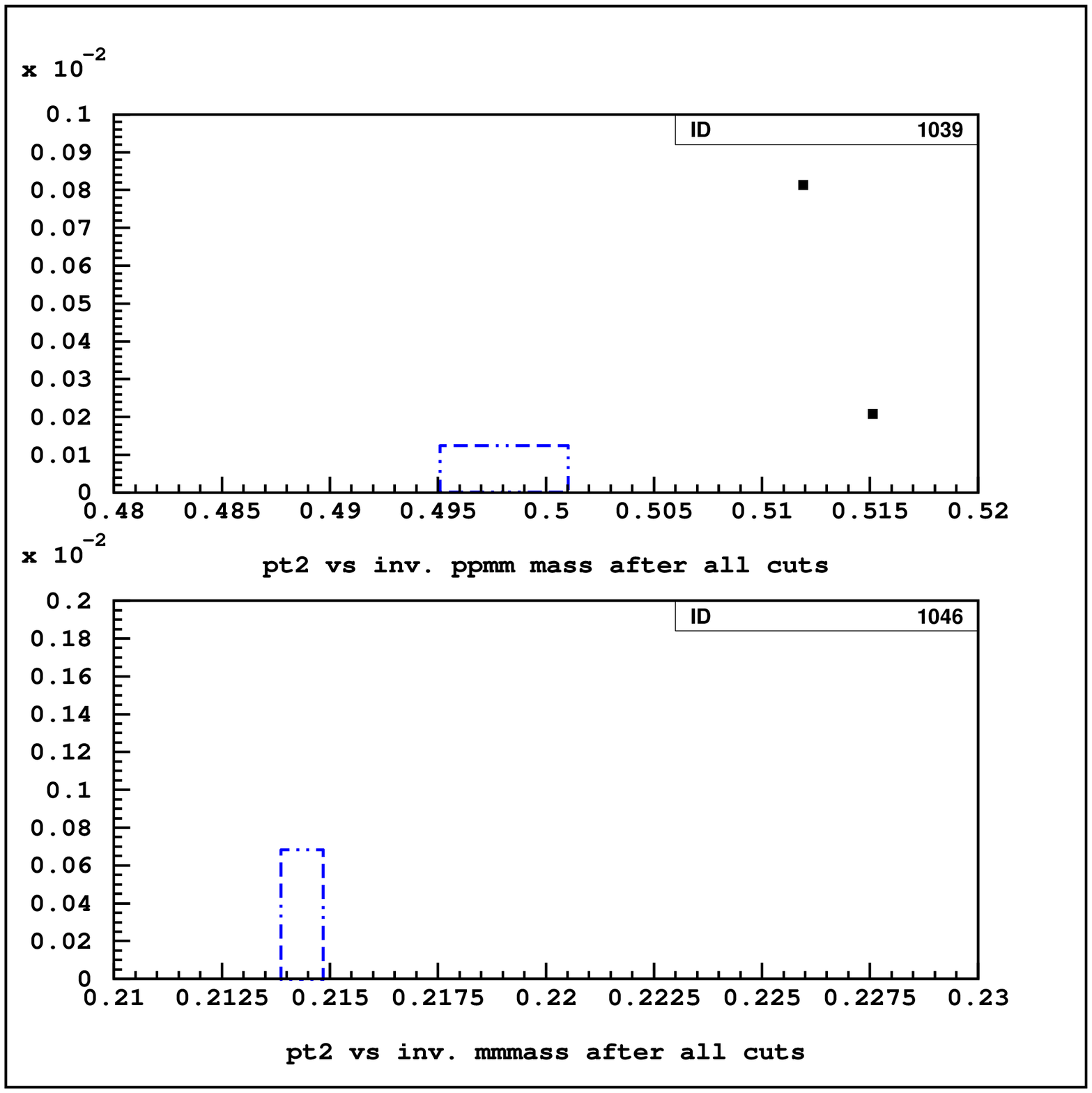}}
    \caption{Left, top: $p_{T}^{2}$ \vs~$M_{\mu\mu\pi\pi}$ for candidate events
in the 1996 data set.  Left, bottom: $|p_{T,\mu\mu}^{2} - p_{T,\pi\pi}^{2}|$
\vs~$M_{\mu\mu}$ for candidate events in the 1996 data set.  Right, top:
$p_{T}^{2}$ \vs~$M_{\mu\mu\pi\pi}$ for candidate events in the 1999 data set.
Right, bottom: $|p_{T,\mu\mu}^{2} - p_{T,\pi\pi}^{2}|$ \vs~$M_{\mu\mu}$ for
candidate events in the 1999 data set.  All plots are shown immediately after
the masked signal boxes were opened, which are indicated by dotted blue boxes.}
\label{Opened}
\end{figure*}

The search was conducted using standard 'blind analysis' techniques, and was
done separately on the data sets taken starting in 1996 and in 1999.
Figure~\ref{Opened} shows the resulting distributions of candidate events
in these two data sets.  There being no events in the signal regions, we use
the methodology of reference~\cite{CnH} to set 90\% C.L. limits of

$$ \mathrm {Br}(\mathrm {K^0_{\mathrm L}} \to \pi^0 \pi^0 X^0,
						 X^0 \to \mu^+ \mu^-) < 9.44 \times 10^{-11}	$$
and
$$ \mathrm {Br}(\mathrm {K^0_{\mathrm L}} \to \pi^0 \pi^0 \mu^+ \mu^-)
						< 8.63 \times 10^{-11}.							$$

This result is some 90 times below the predictions for the pseudoscalar $a$
hypothesis; that hypothesis no longer appears tenable.  Comparison of our
result with the predictions based on an axial vector $a$ are less conclusive,
particularly in light of the large uncertainties on that prediction.  This
hypothesis must still be considered possible.

\bigskip
\begin{acknowledgments}
We thank the Fermi National Accelerator Laboratory staff for their
contributions. This work was supported by the U.S. Department of Energy,
the U.S. National Science Foundation, the Ministry of Education and Science of
Japan, the Fundao de Amparo a Pesquisa do Estado de Sao Paulo-FAPESP, the
Conselho Nacional de Desenvolvimento Cientifico e Tecnologico-CNPq, and the
CAPES-Ministerio da Educao. 

I would also like very much to thank our hard-working conference organizers for
this very productive meeting and for their gracious hospitality.
\end{acknowledgments}

\bigskip 

\end{document}